\documentclass[manuscript]{aastex}
\usepackage{amsmath}

\slugcomment{Not to appear in Nonlearned J., 45.}

\shorttitle{Spicule Study} \shortauthors{Zhang et al.}

\begin{document}

\title{Revision of Solar Spicule Classification}

\author{Y. Z. Zhang\altaffilmark{1,2}, K. Shibata\altaffilmark{2}, J. X. Wang\altaffilmark{1}, X. J.
Mao\altaffilmark{1,3}, T. Matsumoto\altaffilmark{4}, Y.
Liu\altaffilmark{5}, and J. T. Su\altaffilmark{1}}

\altaffiltext{1}{Key Laboratory of Solar Activity, National
Astronomical Observatories, Chinese Academy of Sciences, Beijing
100012, China; yuzong@nao.cas.cn; wangjx@nao.cas.cn; sjt@bao.ac.cn}
\altaffiltext{2}{Kwasan and Hida Observatory, Kyoto University,
Kamitakara, Gifu 506-1314, Japan; shibata@kwasan.kyoto-u.ac.jp}
\altaffiltext{3}{Department of Astronomy, Beijing Normal University,
China; maoxj@bnu.edu.cn}\altaffiltext{4} {Institute of Space and
Astronautical Science, Japan Aerospace Exploration Agency, Japan;
takuma.matsumoto@gmail.com} \altaffiltext{5} {Yunnan Astronomical
Observatory, National Astronomical Observatories, China;
lyu@ynao.ac.cn}

\begin{abstract}
Solar spicules are the fundamental magnetic structures in the
chromosphere and considered to play a key role in channeling the
chromosphere and corona. Recently, it was suggested
by De Pontieu et al. that there were
two types of spicules with very different dynamic properties, which
were detected by space-time plot technique in the $\mbox{Ca {\sc
{ii}}}$ H line (3968 \AA) wavelength from Hinode/SOT observations.
`Type I' spicule, with a 3-7 minute lifetime, undergoes a cycle of
upward and downward motion; in contrast, `Type II' spicule fades away
within dozens of seconds, without descending phase. We are motivated
by the fact that for a spicule with complicated 3D motion, the
space-time plot, which is made through a slit on a fixed position,
could not match the spicule behavior all the time and might lose its real
life story. By revisiting the same data sets, we identify and
trace 105 and 102 spicules in quiet sun (QS) and coronal hole (CH),
respectively, and obtain their statistical dynamic properties.
First, we have not found a single convincing example of `Type II'
spicules. Secondly, more than 60$\%$ of the identified spicules in
each region show a complete cycle, i.e., majority
spicules are `Type I'. Thirdly, the lifetime of spicules in QS and
CH are 148 s and 112 s, respectively, but there is no fundamental
lifetime difference between the spicules in QS and CH 
reported earlier. Therefore, the suggestion of coronal heating by
`Type II' spicules should be taken with cautions.

\end{abstract}


\section{Introduction}
Spicule was initially discovered  by \citet{sec1877} of Vatican
Observatory in Rome
 and later was named as spicule by \citet{rob45}. It has jetlike luminous structure for its
long and slim profile \citep{bec72, lor96}. Generally, spicule could
be seen through chromospheric lines, such as H$\alpha$, H$\beta$, D3
and $\mbox{Ca {\sc {ii}}}$ H and K \citep{mic54}. In recent decades,
larger size spicules, with a similar structure, were observed in
$\mbox{He {\sc {ii}}}$, ultraviolet (UV), extreme-ultraviolet (EUV)
and soft X-ray wavelengths, called UV, EUV or soft X-ray
macrospicules, respectively \citep{boh75, der89, wil00, xia05}.

A huge number of spicules, like messy hair, and the inter-spicule are
composed of the chromosphere. The mass flux taken
 by spicules to corona is exceeding that by solar wind by two order of
 magnitude \citep{tho61}. Therefore, besides working as the 
indication of inhomogenous chromosphere, spicule is thought to be
a very likely candidate in transporting the material and kinetic
energy into the corona as well as heating the corona
\citep{wol54,rus54,li09}. For its mysterious formation mechanism and
possibility in heating the corona, all the time, it attracts
researchers' strong attentions and interests. \citet{sue82}
suggested that spicules are formed as a result of slow mode shocks
propagating along vertical magnetic flux tubes in the photosphere
and low chromosphere using one dimensional hydrodynamic simulation.
\citet{shi82} explained why spicules are taller in coronal hole (CH)
(Lippincott 1957, Beckers 1968, 1972) by extending the slow shock
model. Later the Alfv$\acute{e}$n wave model was successfully
proposed to explain spicules and their role in heating corona
\citep{hol82, ste88, hol92, kud99}. It should, however, be noted
that even in the Alfv$\acute{e}$n wave model slow shocks are
generated due to nonlinear mode coupling with Alfv$\acute{e}$n waves
and play an essential role in the acceleration of spicules
\citep{sai01}. \citet{pon07b} discovered ubiquitous Alfv$\acute{e}$n
waves on spicules by Hinode/SOT. Ubiquitous horizontal field
discovered in the photosphere \citep{lit08,ish08, jin09,zha09} might
trigger reconnection in the photosphere and low chromosphere which
were suggested to excite Alfv$\acute{e}$n waves \citep{tak01, iso08}.
Based on the model proposed by \citet{suz05,suz06}, by considering
observed photospheric granular buffeting as the source of
Alfv$\acute{e}$n waves, \citet{mat10} successfully reproduced
spicules, corona and solar wind.

Previous observations of individual spicule were 
difficult because of the
low observation resolution \citep{ste00}. The situation has been
much improved since the built of Swedish 1 m Solar Telescope (SST)
\citep{sch03} in 2003 and the launch of Hinode satellite
\citep{tsu08, sue08, ich08} in 2006.
Hence, it is meaningful to
re-measure the dynamic properties of spicules with these seeing free
data sets and to further study the coronal heating.
According to the new observations, \citet{pon07a} claimed that the
spicules should be divided into two types: `Type I' spicule with 3-7
minute lifetime is driven by shock waves that are formed as a result of
p-mode leakage; and `Type II', a result of magnetic reconnection,
bears much shorter lifetime about 10-150 seconds and faster speed
between 50-150 km $s^{-1}$. The `Type II' spicules dominate the
structure of solar chromosphere in CH. Without a downward moving
phase, `Type II' spicules fade away promptly in the corona \citep{pon07a,
ste10}, so it seemed natural to accept this mechanism to explain the
coronal heating.

Yet, from the filtergrams, the movements of spicules could be
observed in both horizontal and vertical directions. If spectral
observations were available, we could get the Doppler shift in the
line of sight \citep{nik67, pas68, wea70, sue95}. Spicule movement
usually appears in a more complicated 3D way, but we don't know what its real 
movement is. The movements observed in filtergrams are
thought to be related to the waves, such as kink waves and
Alfv$\acute{e}$n waves \citep{pon07b}, and oscillations
\citep{kul83}. We notice that the space-time plot, which could
reflect the dynamic information of spicule in a certain extent
\citep{ban00, chr01, tav11}, was applied directly to measure the
lifetime and height of a spicule in their study. For the real 3D
motion of a spicule as mentioned above, especially the extensive
lateral movement, a spicule could not always keep its
motion in a fixed direction. Therefore, some doubts arise on the
reliability of the lifetime and height measured by the method.

We are motivated, therefore, to re-measure the dynamical properties
for a better understanding of the spicule model. In this work the
data sets in \citet{pon07a} are revisited. We first re-examine a few
types of morphology that was regarded as spicule by
\citet{pon07a}. Are they real spicules? If not, what is the
distinction between them and those spicules observed through filtergrams
directly? By drawing a comparison on the height and lifetime, it is
discovered that the `spicules', identified in space-time plots, usually own
shorter lifetime and lower height than those in
filtergrams. To show the statistical properties of spicules then 
we trace 105 and 102 spicules in QS and CH, respectively
and survey their distributions of lifetime, height, velocity and
acceleration. 

Section 2 will introduce the observations and method adopted in this
paper. The results will be listed in Section 3. The discussion and conclusion 
are in the last section.

\section{Observations and Analysis}

\subsection{Data}
 We use the same data sets in QS and in CH adopted by \citet{pon07a}.
The observations were carried out by the Hinode/SOT Broadband Filter
Imager (BFI) in $\mbox{Ca {\sc {ii}}}$ H filter whose bandwidth is
broad enough to observe both photosphere and chromosphere
simultaneously.

The selected QS, near to an $\alpha$ \sbond type active region, NOAA
10923, is located in the western limb of the Sun with the center
coordinates of 960 $\arcsec$, -90 $\arcsec$ and the field of view
(FOV) of 56 $\arcsec$ $\times$ 56 $\arcsec$. From 00:00:04 UT to
02:19:59 UT on 2006 November 22, there are totally 1,050 frames
observed with a pixel size of 0.05 $\arcsec$ and time cadence of
about eight seconds. The center coordinates of the CH are of 0
$\arcsec$, -968 $\arcsec$ with the same size of FOV as the QS. Totally
758 frames are available in observations from 11:29:32 UT to
12:30:00 UT on 2007 March 19. The spatial resolution remained the
same but the temporal resolution was improved to five seconds or so.
The IDL routine in the libraries of Solar Software, fg$\_$prep.pro
is applied to the image reduction to correct dark currents and other
errors of the camera. Then we remove the cumulative offsets and have
the data set co-aligned. Additionally, we use the radial density filter \citep{oka07} to 
enhance the visibility of spicules.

\subsection{Comparison of Spicules Observed by Two Methods}

In the space-time diagrams, totally five typical morphology appearance
of `spicules' have been identified with large discrepancies of their lifetime and movement mode 
as classified by De Pontieu et al. (2007a). The `Type I' spicules are indicated by `A' and `B'
in Figure 1 and the `Type II' spicules by `C' and `D' in Figure 1 and by `E' in Figure 2.
The tops denoted by triangles in Figure 1 and 2 are determined by
the space-time plot at seven points of time for each case.

Yet, it is uncertain whether the ture height is
obtained or not. To make sure of it, we try to find the tops of the
`spicules' in the filtergrams. In Figure 1, starting with the third
panel, its seven panels in the row are the
filtergrams corresponding to the seven points of time. The plus
indicates the true height
 of the spicule directly determined by the filtergrams.
The scatter plots in Figure 1 and 2 are used to compare the
positions of the tops acquired by two methods, respectively.
 For example, in the panel marked with `A' in Figure 1, the plus signs
 basically match those triangles. It means in this case the lifetime and height
determined by space-time plot is basically true. However, in the
case of `B', the triangles are always slightly lower than
those pluses though only several pixel distance apart
between the slit for the space-time plot in the example `B' and
the one in the example `A'. The errors of the height
measured in the way of space-time plot would produce errors in 
velocity and acceleration as well. In the sample `C' we find that there
are also great discrepancy in height measurement between
two methods. In the space-time plot, the `spicule' seems falling
off suddenly, however, according to the filtergrams, the spicule
falls more slowly and does not finish its whole life experience as
shown in the corresponding scatter plots. The example `D'
is a more `typical' `Type II' spicule in space-time plot, but in the
filtergram method the undetected descent phase
by the first method did exist. As to the last example `E'
(shown in Figure 2) having a strange shape, the reason of lacking
the ascent phase is that the first half process was not yet recorded
in the space-time plot. The entire lifetime should include the panels
enclosed by the dotted frame indicating the ascent phase and
the dashed frame, the descent phase, but the
space-time plot failed to detect the first part. Morphologically,
in the space-time plots, as having been pointed out by
\citet{pon07a}, two types of spicules are identified. But the
approximately vertical stripes (`Type II' spicule) may be caused by
the employment of the slit, which just cuts a part of the brightness
distribution of a spicule, so that only a stripe of the distribution
is left to be shown in the plots. For this reason, we hope to get statistical
result of the spicule dynamic properties to answer if there are two types of spicules
and their distributions in QS and CH.

\subsection{The Measurement of the Kinetic Parameters of Spicule}
In the data set of QS, we identify 36, 33 and 36 spicules in three
frames, i.e., Frames No. 100, 300 and 500 of total 1,050 filtergrams
observed by $\mbox{Ca {\sc {ii}}}$ H filter, respectively. 
The upper panel of Figure 3 shows the No. 300 frame shot in QS at
00:40:05 UT on 2006 November 22. The tops of the identified
33 spicules (numbered 37 to 69) in this frame are marked by small
squares in five color $\sbond$ red, green, blue, purple and yellow
in turn, in the meanwhile their serial numbers are also written in
the same color over the indicated spicules. The white
dashed squares illustrate the FOVs of those filtergrams in
Figures 4 and 5, respectively.

Figure 4 represents the time series of the dynamic process of No. 55
spicule in QS (hereafter called `SQ 55'). The spicule,
as shown from the third panel to the third last panel, is
 detected from 00:39:57 UT to 00:43:17 UT. To find the top of the
spicule at each time, based on the judgement with naked
eyes, we used a set of brightness curves as an auxiliary
measure to acquire more precise position for each top. These
brightness curves reflect the brightness variation along the 13
horizontal lines with a two-pixel distance separating two
adjacent lines. In case of these lines covering the spicule SQ 55,
\textbf{these} brightness curves are plotted exactly above the
spicule. Among them, the red brightness curve displays the
brightness variation of the red dotted horizontal line. 
A spine shape is being gradually formed,
becoming quite evident at about 00:41:09 UT, and then weakening
slowly. At 00:43:17 UT, it was almost undetectable for its dropping 
to a lower position and then mixing with other spicules. The upper end of the spine,
which is almost as weak as its ambient, should be the position
 of the top of the spicule at each
moment. A series of red triangles corresponding to the time 
are used to indicate the spicule
top positions. A green dotted vertical line
and its concolorous brightness curve are plotted just as a
reference. The brightness has an apparent variation around the
top. A long red dashed vertical line marks the
brightness value of 20 in vertical direction. In the bottom right
panel in Figure 4, is the height-time plot, and the spicule 
height at each moment determined by the
filtergrams is shown with the diamond. The apparent lifetime and
height of the spicule SQ 55 are 200 seconds and 4,819 kilometers,
respectively. The top trajectory marked by the diamonds shows a
typical parabolic profile covering a complete cycle of acsent and
descent. Before 00:39:57 UT and after 00:43:17 UT, due to the 
overlapping by a very large
number of spicules, the SQ 55 could not be detected exactly. 
Evidently, the real lifetime should be
longer than the apparent one, though we could not tell the real path
of the spicule in which it was overlapped by other
spicules. Totally, as shown by Table 1, there are 
71 of 105 (67.6$\%$) spicules in QS
moving upward firstly and then falling back. As for other 34 spicules, as shown in Figure 6,
it seems that there are three different `types' of spicules. The first `type' only has the ascent phase, indicating
by SQ 82; the second as SQ 33 just descending from higher to lower; and the last one, denoted by SQ 44, shows 
no obvious ascending and descending behaviors. Thus the profiles of these three types are likely the result of trajectory
mixture with surrouding spicules. We still could not trace the whole life for individual.

In the data set of CH, similarly, from No. 150, 300, 450 and 600
frames, 20, 24, 34 and 24 spicules have been identified and traced
, respectively. In the lower panel of Figure 3 is the No.
150 frame with 20 identified spicules indicated with the same
symbols as above. The added white dashed frame is similar to the
illustration of the FOV used in the filtergrams of Figure 5. Figure
5 shows the whole tracing process of No. 17 spicule in the coronal
hole (hereafter called `SC 17') from 11:40:43 UT to 11:44:38 UT, 
i.e. its apparent lifetime being
235 seconds. In contrast to the spicules in QS, the peaks
of spicules in CH are too dim to be accurately detected. In the left lower panel of
Figure 5, the tops at each time determined by frame tracing are
shown in the height-time diagram. Apparently, SC 17 also
experienced a complete cycle, up and then down. As listed by the right column of Table 1, 
total 102 spicules in CH, 64 (62.7$\%$), 24 (23.5$\%$), 
6 (5.9$\%$) have been acquired the up- and downward, only upward, only downward phases; the remaining 8
(7.9$\%$) spicules without apparent upward or downward phase.

\subsection{The Dynamic Properties of Spicules in QS and CH Regions}
By tracing 105 and 102 spicule in QS and CH, we obtain the
histograms of height, lifetime and vertical velocity shown in Figure
7. In panels A and B, the mean height in QS is 5,014 km ranged
from 1,027 to 8,690 km, which is much lower than that, 9,592 km,
from 4,819 to 17,142 km in CH. The average mean apparent lifetime in
QS is 148 s, while in CH, it is 112 s,
respectively. According to the height-time plot of each spicule in
QS and in CH, we find that there are 71 of total 105 (about
67.6$\%$) spicules in QS and 64 of total 102 (about
62.7$\%$) in CH having a relatively complete cycle of ascent
and descent. The reason of those spicules without a cycle basically
lies in the disorder overlapping of a huge number of spicules. Hence
for some spicules we observed the whole process of rising from the
back of their adjacent spicules and then falling down; and for
others only observed ascending, or descending or intermediate
stage between. Therefore the measured lifetime should be shorter than their real lifetime.
Yet, no one could tell what will occur when a spicule submerges 
in its background. Additionally, the
velocity is basically proportional to the deceleration both in QS and in CH
(see panels G and H in Figure 7).

\section{Discussion and Conclusion}

For the lateral motion of a spicule either in QS or in CH, no fixed
slit could be employed in its whole lifetime.
Therefore the spicule profile formed from space-time plot could not
exactly represent the real trajectory of its movement. Comparing the
kinematic parameters of spicules obtained in both methods, the
height, lifetime acquired with the first method are always much less
than those with the method we have adopted. That is why we do
not use the space-time plot but employ its filtergrams directly to
trace the trajectory of each spicule. Both in QS and in CH, 
are more than 60$\%$ spicules moving in a complete cycle of ascent and descent, 
and the rest showing no cycle suggested to be mixed
in the background. In brief, with the same data sets in QS and in CH
already used by \citet{pon07a}, no convincing `Type II' spicule
has been captured. Therefore, `Type I', but not `Type II' spicule dominates in QS and
CH as shown in Table 1. \textbf{This result} is consistent with the study of spicules in
the disk over QS \citep{sue98} and over a plage \citep{ana10} and at
the limb \citep{pas09}. \textbf{It suggests that there is no a sufficient number of `Type II' spicule to heat corona by their fading way both in QS and in CH.}

Surely, there are apparent discrepancies
in the dynamic properties of spicules in QS and in CH. The spicules
in CH seem more energetic than those in QS \citep{shi82}, for instance,
they hit much higher position with higher speed and their lifetime
is shorter. However, the relation between velocity and deceleration
both in QS and in CH is approximately directly proportional associated with
some different coefficient. There is a kind of faster spicules, but
this does not mean that their physical mechanism is essentially different from
that of other spicules. 

\textbf{With the help of the unprecedentedly high
spatial resolution of Hinode/SOT $\mbox{Ca {\sc {ii}}}$ H filter,
the identified diameter of spicule could be less than 200 km, which
is much thinner than the values by observations in the past.
However,} it is still not easy to acquire
a complete image of a limb spicule. Because of the faintness at the top of
the spicule, there is some uncertainty in determining the top of a
spicule \citep{rus54}. At the lower position, 
as \citet{lyn73} pointed out, the
spicules are so crowded, only when the visible spicules rise to
some height and separate from each other far apart we could measure their kinetic parameters
\citep{wol54}. It means that the measurement of the dynamic
properties of spicules is far from being perfect. Thus the life story
of these spicules remains somewhat vague. New instruments
and technique are reasonably expected.

\textbf{Last but not the least}, when we watch the animation of $\mbox{Ca {\sc {ii}}}$ H
filtergrams, spicules could form a `group', moving in a similar
way, such as dancing a waltz, or as a bamboo raft dispersing into
individuals, etc. The `group' motion may behave a more
complicated behavior of a double thread structure of spicule having
the following evolution (expansion thread separation, lateral motion
and spinning as a whole body) \citep{sue08a, ste10}, which was
speculated as magnetic reconnection. Thus, besides the energy taken
by spicules themselves, these intensive activities will carry a huge
amount of energy from chromosphere into corona. It may be the
essentials of macrospicule and the practical way of releasing energy to
corona. Our next work will pay more attention to measuring the
kinematic parameters in 3D combining with ground-based Doppler
observations in the line of sight \citep{sho10}.

\acknowledgments

Hinode is a Japanese mission developed and launched by ISAS/JAXA,
collaborating with NAOJ as a domestic partner, NASA and STFC (UK) as
international partners. Scientific operation of the Hinode mission
is conducted by the Hinode science team organized at ISAS/JAXA. This
team mainly consists of scientists from institutes in the partner
countries. Support for the post-launch operation is provided by JAXA
and NAOJ (Japan), STFC (U.K.), NASA, ESA, and NSC (Norway).
 We are grateful to our anonymous referee
for his/her insightful comments and suggestions helped us to improve the manuscript considerably. The
discussion with Dr. H. Isobe, Dr. P. F. Chen, H. Watanabe, T.
Kawate, T. Anan is very appreciated. This work is supported by the
Young Researcher Grant of National Astronomical Observatories,
Chinese Academy of Sciences (118900KX3), the National Natural
Science Foundations of China (40890161, 40974112, 11025315, and 11003026), the
CAS Project (KJCX2-EW-T07), and the National Key 
Basic Research Science Foundation Program of China (2011CB811403) and Y07024A900.

\clearpage

\begin{table*}
  \caption{The statistical dynamic properties of four `types' of spicules}\label{tab:1}
\small
  \begin{tabular}{c | c c ccc | c ccc c}     
  \hline
  \hline
& & & QS & & &&& CH &&\\

\hline
Properties & Num. & H & LF & Vy & a &  Num. & H  & LF & Vy  & a \\
&  & (km) & (s) & (km s$^{-1}$) & (km s$^{-2}$) &   & (km) & (s) & (km s$^{-1}$) & (km s$^{-2}$) \\
\hline
Up- and Downward & 71 & 5174 & 176 & 16.9 & -0.13 & 64 & 9572 & 121 & 48.0 & -1.37 \\
Only Upward & 9 & 4792 & 74 & 11.2 & -0.10 & 24 & 10,391 & 100 & 26.7 & -0.22 \\
Only Downward & 11 & 4729 & 82 & -9.6 & 0.05 & 6 & 8176 & 86 & -6.7 & -0.39 \\
Uncertainty & 14 & 4570 & 103 & - & - & 8 & 8418 & 95 & - & - \\
Sum(Mean) & 105 & 5014 & 148 & 15.5 & -0.14 & 102 & 9592 & 112 & 40.5 & -1.04 \\

\hline
\end{tabular}
{Notes--`Num.', `H', `LF', `Vy', and `a' means the number, mean height, mean lifetime, mean velocity in vertial direction and acceleration in vertial direction, respectively.}
\end{table*}
\clearpage

\begin{figure}
\epsscale{0.95} \plotone{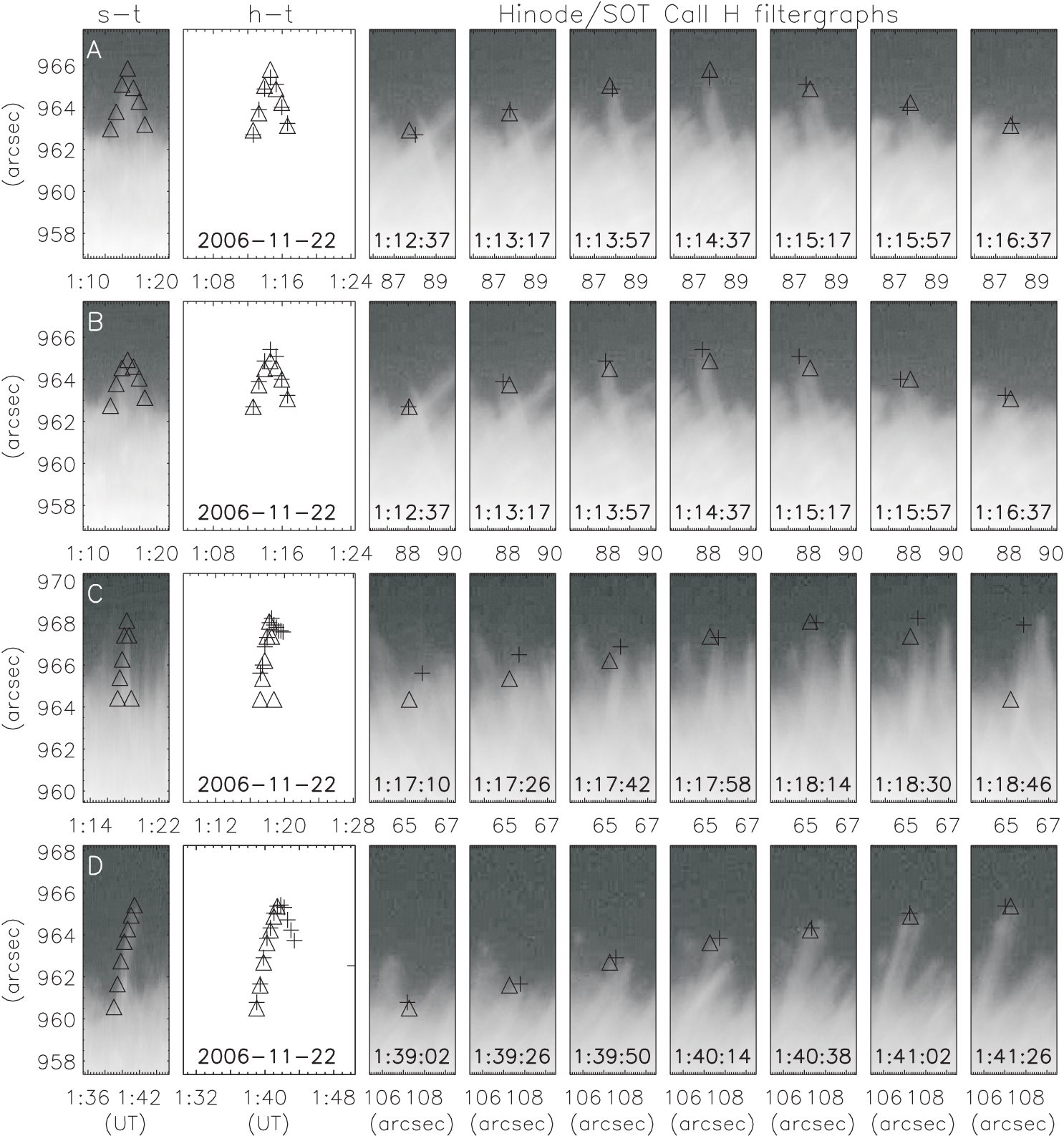}\caption{Comparison of two
methods for tracking spicules. In the panels of the first column are
four typical morphology appearance of `spicules' determined by space-time plots; and the second
column are the height-time plots. The last seven columns are
$\mbox{Ca {\sc {ii}}}$ H filtergrams observed by Hinode/SOT. The
triangle denotes the height acquired by space-time plot; as a comparison, the plus is
the height obtained by frame tracking promptly. \label{fig1}}
\end{figure}
\clearpage

\begin{figure}
\epsscale{0.95} \plotone{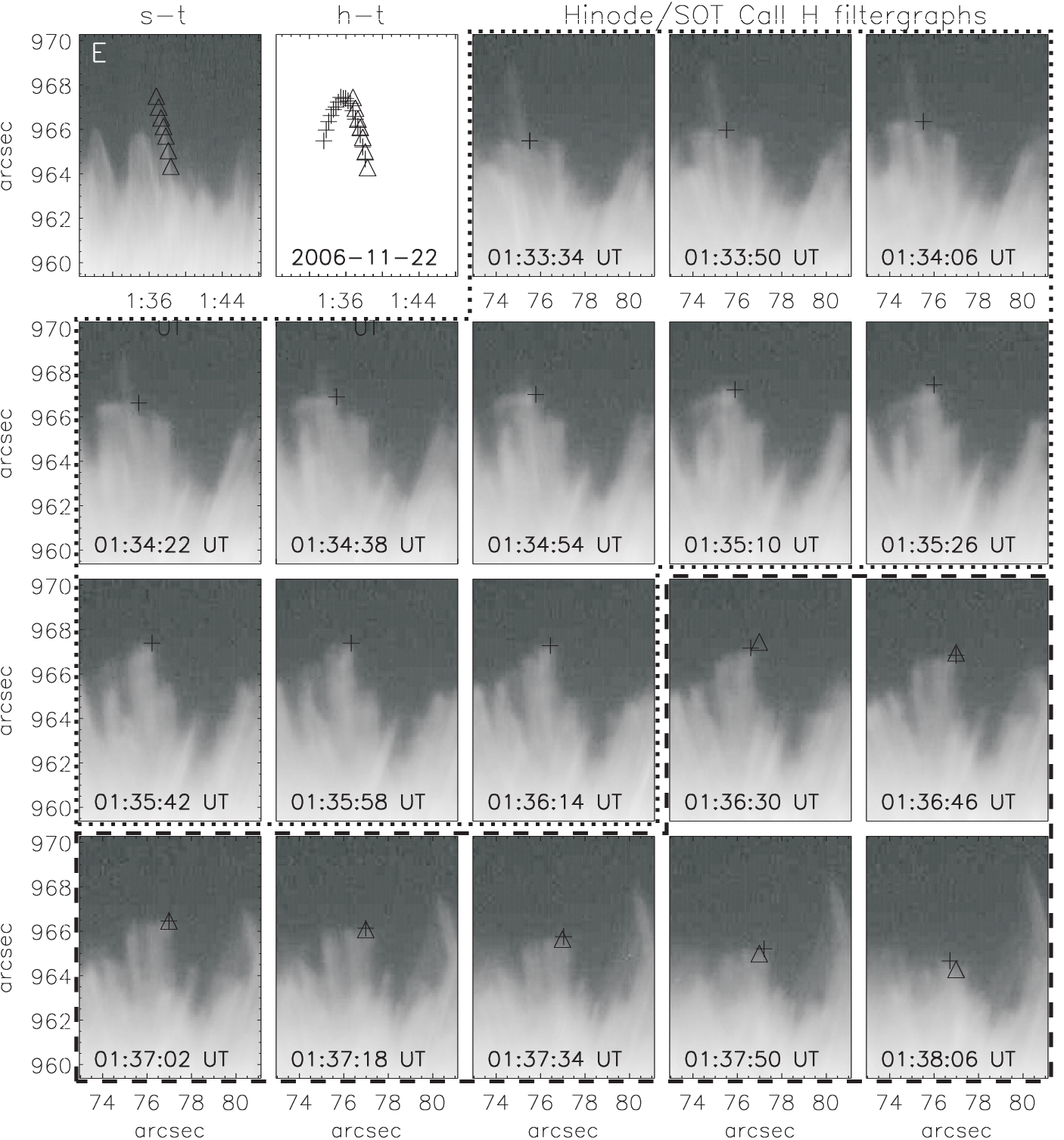}\caption{Same as Figure 1
but for the fifth typical morphology appearance of `spicule'. The first panel in the top left shows the `spicule'
identified by the space-time plot; the second panel is then the height-
time plots; from the third panel to the last one in the bottom right are the $\mbox{Ca
{\sc {ii}}}$ H filtergrams observed by Hinode/SOT. The triangles and
the pluses are the same as Figure 1. \label{fig2}}
\end{figure}
\clearpage

\begin{figure}
\epsscale{.95} \plotone{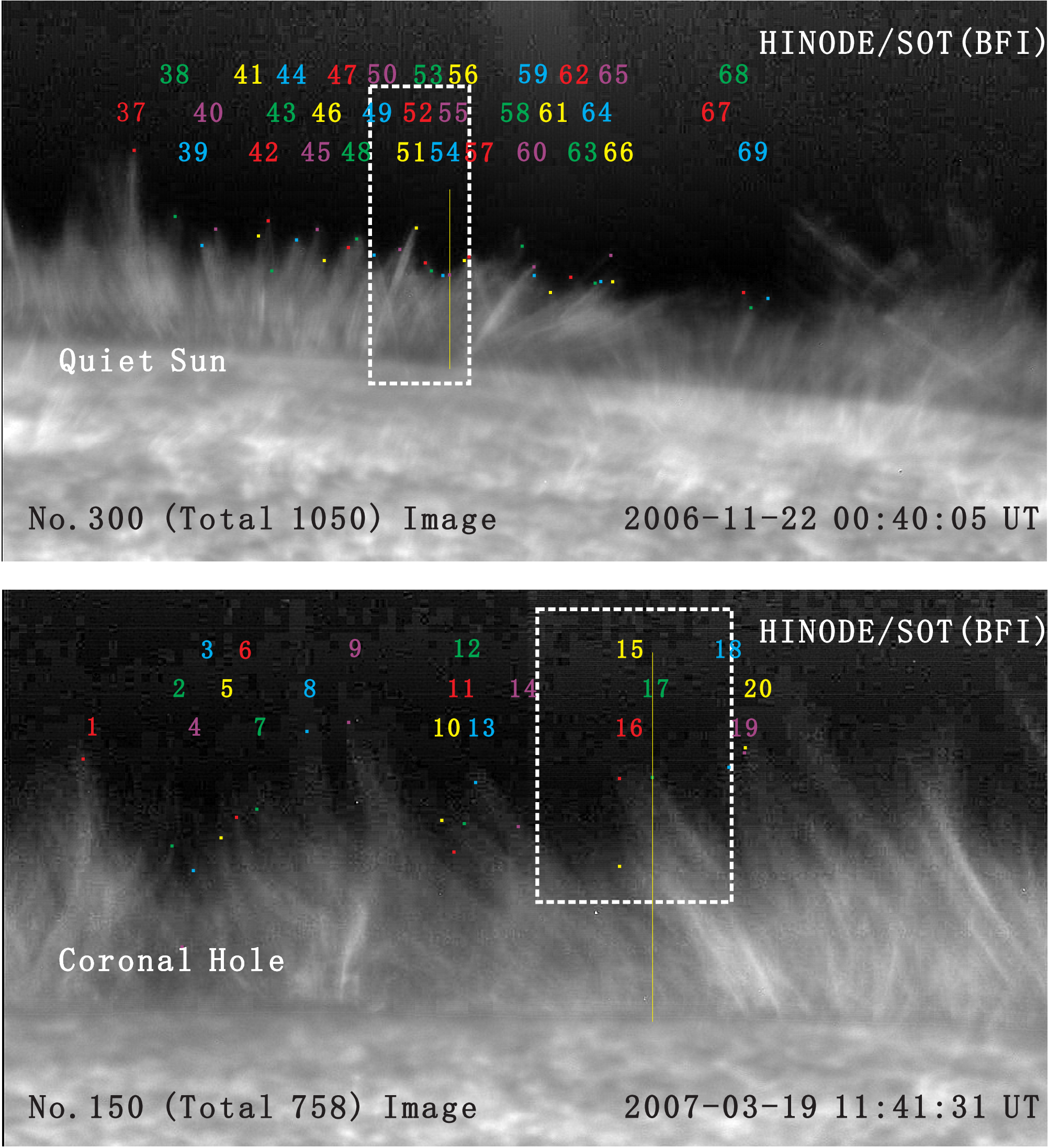}\caption{$\mbox{Ca{\sc
{ii}}}$ H images acquired by Hinode/SOT for quiet
Sun at 00:40:05 UT on 2006 November 22 (upper panel) and coronal
hole at 11:41:31 UT on 2007 March 19 (lower panel). Those colorful
mini-squares marked the top\textbf{s} of identified spicule\textbf{s} numbered
with the concolorous Arabic numerals. The yellow lines denote the
positions of slits to make space-time plots; white dashed frames are
the field of view of filtergrams in Figure 4 and 5, respectively. \label{fig3}}
\end{figure}
\clearpage

\begin{figure}
\epsscale{0.95} \plotone{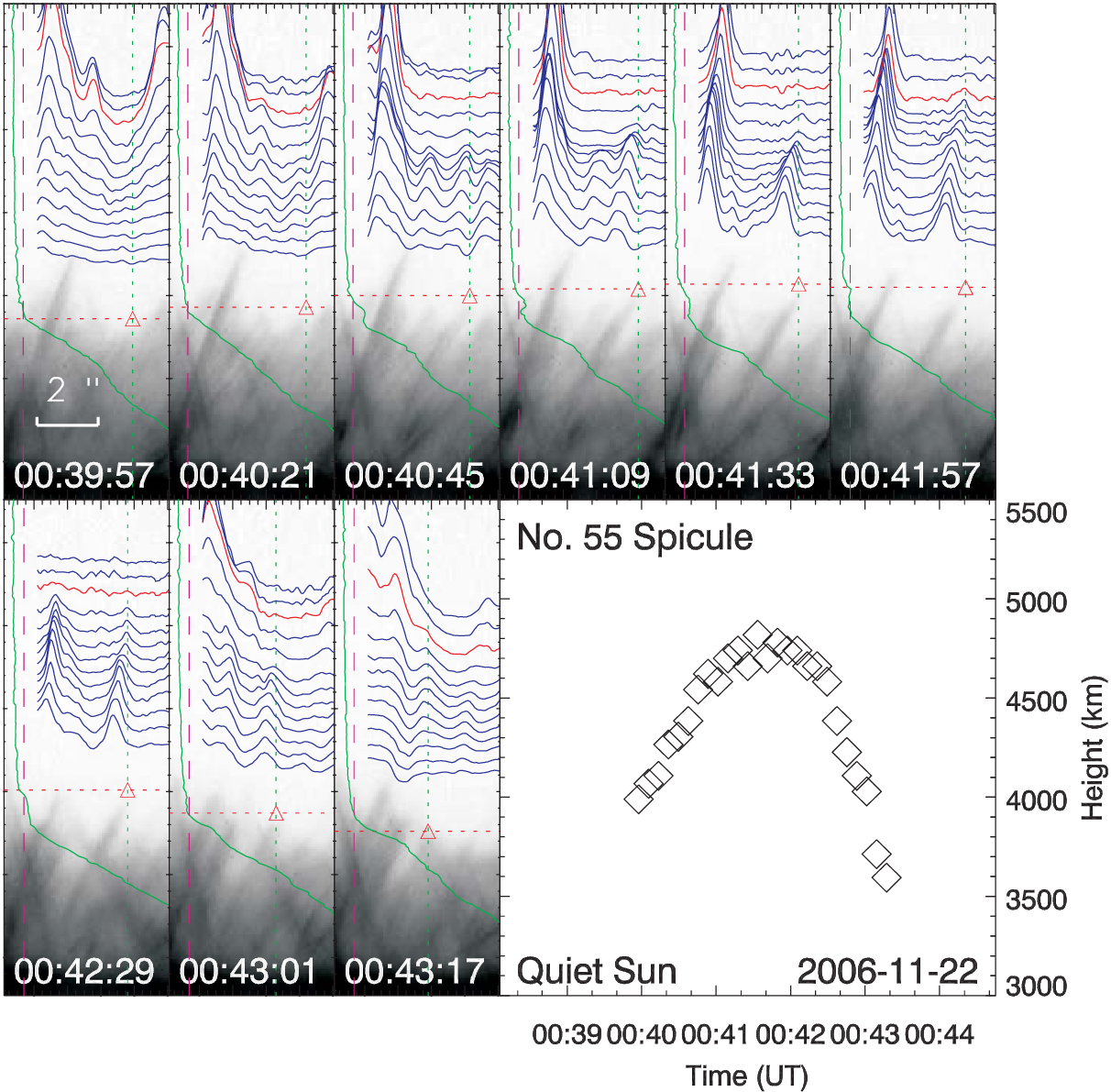}\caption{Time series of
the dynamic processing of No. 55 SQ observed by Hinode/SOT Broadband
Filter Imager in the $\mbox{Ca {\sc {ii}}}$ H on 2006 November 22.
In each sub-images, red triangle marks the top of the spicule.
The green solid curve is the brightness variation along the green
dotted line. The red dashed line denotes the brightness value of
20. Similarly, the red solid curve is the brightness variation along
the red dotted line. Two upper and ten lower cyan solid curves are
also brightness variations along corresponding lines which are parallel to the red dotted line, but upper
and lower than the red dotted line, respectively. The panel at the bottom right is the height-time 
plot of this spicule.  \label{fig4}}
\end{figure}
\clearpage

\begin{figure}
\epsscale{.95} \plotone{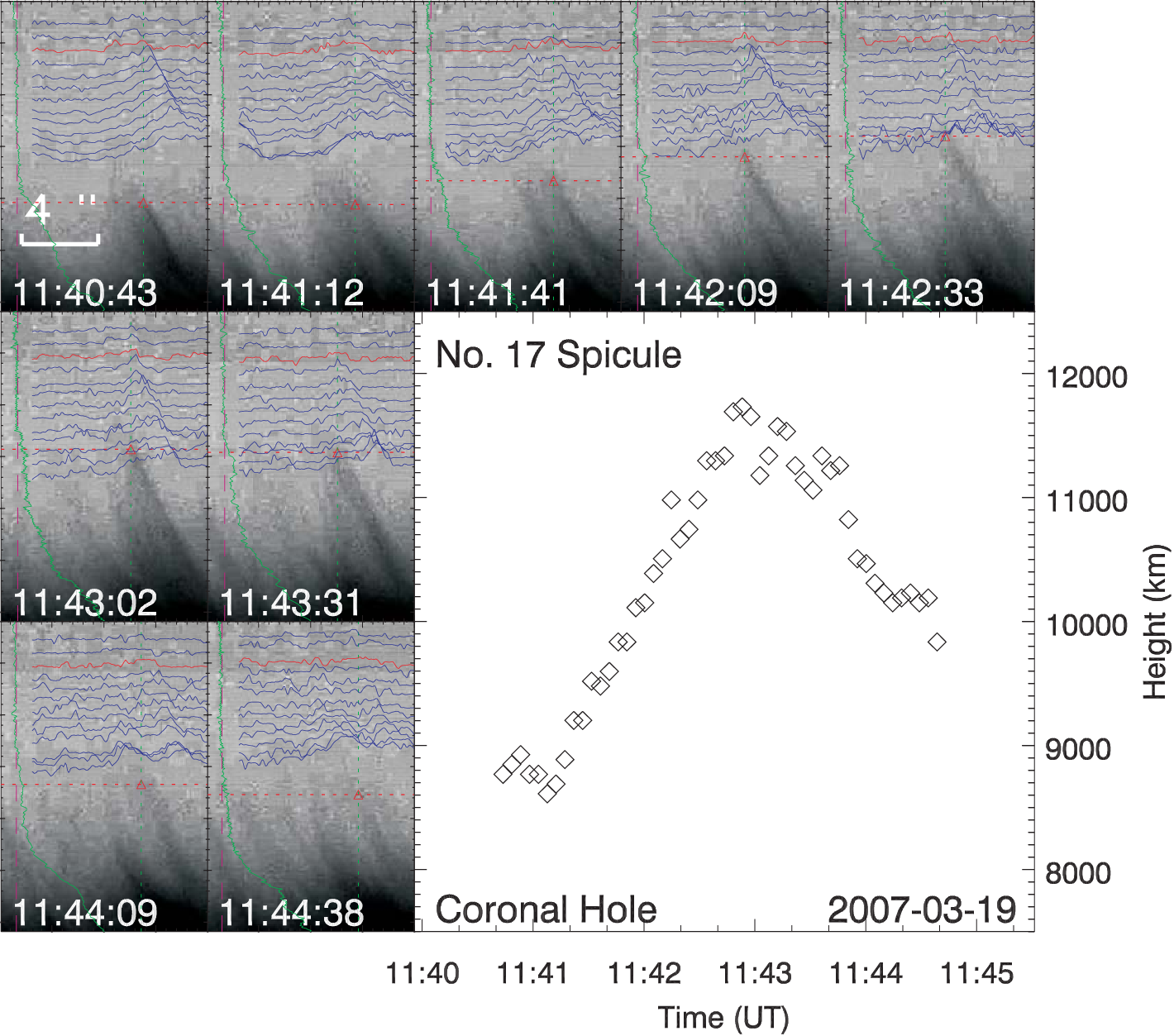}
\caption{Time series of the dynamic process of No. 17 SC observed by Hinode/SOT Broadband Filter Imager in the $\mbox{Ca {\sc {ii}}}$ H on 2007 March 19. The lines
have the same meaning as those described in Figure 4. The bottom right panel is the height-time plot of the spicule, too.\label{fig5}}
\end{figure}
\clearpage

\begin{figure}
\epsscale{0.95} \plotone{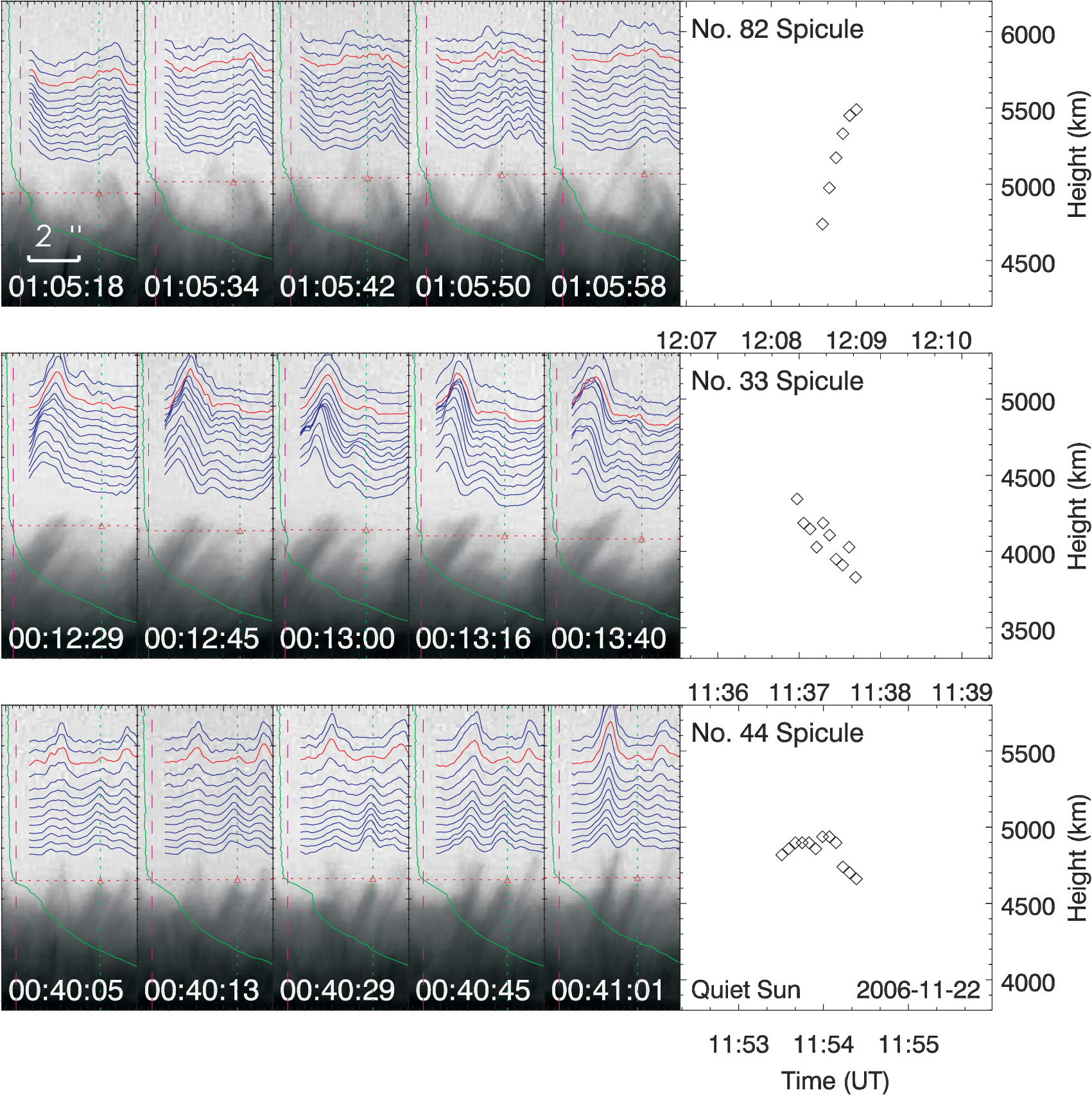}\caption{Time series of the dynamic 
process of other three `types' of spicule and the related height-time plots. 
No. 82 is only observed at the ascent stage; No. 33 at the descent, but No. 44 spicule shows
 no obvious change in its height.\label{fig7}}
\end{figure}
\clearpage

\begin{figure}
\epsscale{0.7} \plotone{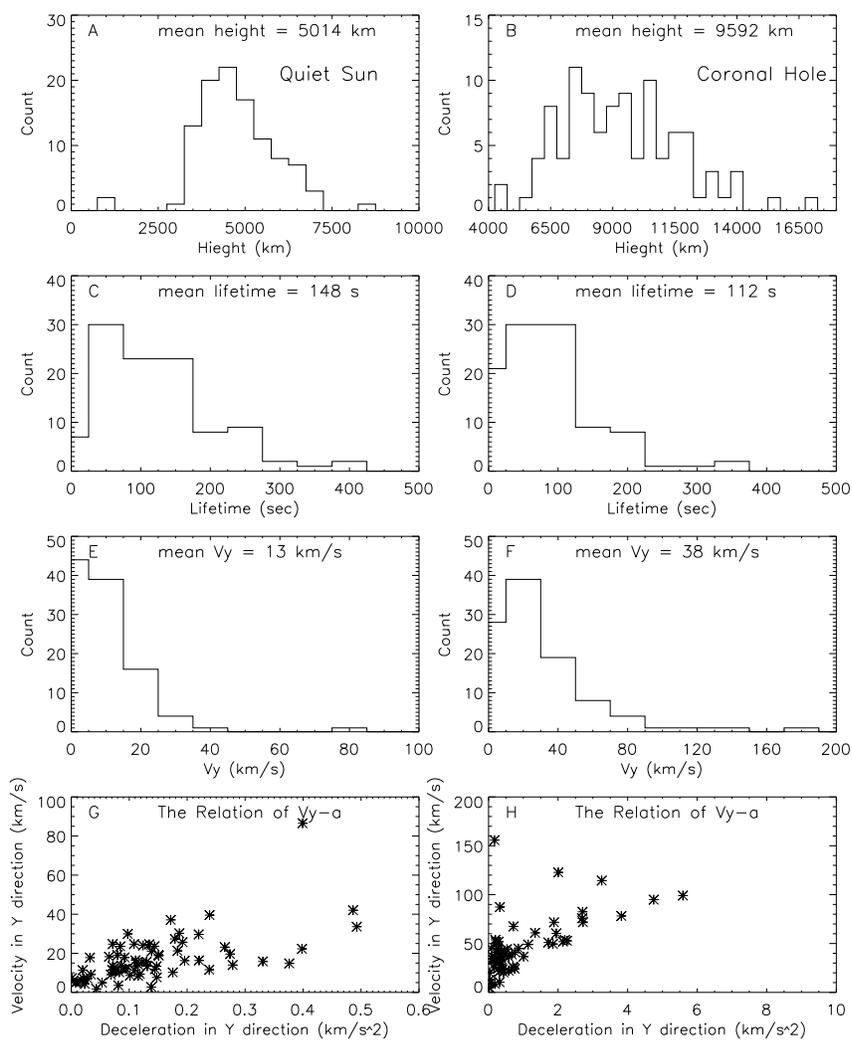} \caption{Histograms for QS (2006 November 22) and CH (2007 March 19). (A)
and (B) the maximum height, (C) and (D) the lifetime, (E) and (F)
the maximum vertical velocity. (G) and (H) are the scatter plot of spicule's maximum
velocity vs. its maximum deceleration in vertical direction.\label{fig8}}
\end{figure}
\clearpage

\end{document}